\documentclass[10pt,letterpaper]{article}
\usepackage{opex3}

\usepackage{graphicx}% Include figure files 
\usepackage[latin1]{inputenc} 
\usepackage{dcolumn} 
\usepackage{hyperref}
\usepackage{xcolor}
\usepackage{mathrsfs}
\usepackage{bm}
\usepackage{amsmath}
\usepackage{amssymb}

\usepackage{soul}
 
\begin{document} 
 
\newcommand*{\DP}[2]{\frac{\partial#1}{\partial#2}} 
\newcommand*{\DPZ}[3]{\left.\frac{\partial#1}{\partial#2}\!\right|_{#3}} 
\newcommand*{\e}[0]{\mathrm{e}} 
\newcommand*{\ii}[0]{\mathrm{i}} 
\newcommand*{\dd}[0]{\mathrm{d}}
\newcommand*{\E}[0]{\mathscr{E}}
\newcommand*{\vect}[1]{\overrightarrow{#1}}
\newcommand*{\omoy}[0]{\omega_0}
\newcommand*{\ovar}[0]{\Delta\omega}
\newcommand*{\GVD}[0]{\mathrm{GVD}}
\newcommand*{\shot}[0]{\mathrm{shot}}
\newcommand*{\K}[0]{{K}}

\newcommand*{\OLR}{\color{purple} }
\newcommand*{\OLC}{\color{teal} }
\newcommand*{\OLN}{\color{olive} }

\newcommand*{\OLRbis}{\color{purple} }
\newcommand*{\OLCbis}{\color{teal} }
\newcommand*{\OLNbis}{\color{olive} }

\newcommand*{\corr}{\color{red}}

\title{Real-time distance measurement immune from atmospheric parameters using optical frequency combs} 

\author{Pu Jian$^1$ , Olivier Pinel$^{1,2}$, Claude Fabre$^1$, Brahim Lamine$^1$ and Nicolas Treps$^{1,*}$}

\address{
$^1$ Laboratoire Kastler Brossel, Université
  Pierre et Marie Curie--Paris 6, ENS, CNRS; 4 place Jussieu, 75252
  Paris, France
\\
$^2$ Centre for Quantum Computation and Communication Technology,
Department of Quantum Science, The Australian National University, Canberra, ACT 0200, Australia
}
 
\email{$^*$nicolas.treps@upmc.fr}

\begin{abstract} We propose a direct and real-time ranging scheme using an optical frequency combs, able to compensate optically for index of refraction variations due to atmospheric parameters. This scheme could be useful for applications requiring stringent precision over a long distance in air, a situation where dispersion becomes the main limitation. The key ingredient is the use of a mode-locked laser as a precise source for multi-wavelength interferometry in a homodyne detection scheme. By shaping temporally the local oscillator, one can directly access the desired parameter (distance) while being insensitive to fluctuations induced by parameters of the environment such as pressure, temperature, humidity and CO$_2$ content.  \end{abstract}
 

The accuracy of precise length measurements is commonly limited by dispersive effects. For instance, the dispersion of air is a crucial issue for geodetic surveying~\cite{Weiss:2001} or for the optical link between a ground station and a satellite~\cite{Exertier:2006}. The lack of knowledge about atmospheric parameters can then act as the main limitation to optical length measurements.

Numerous groups around the world tackle this issue of long range distance measurement.  For instance, in atmospheric links such as satellite ranging or Lunar Laser Range, the accuracy provided by time-of-flight measurements reaches the millimeter level or below~\cite{Exertier:2006,Djerroud:2010}. Interferometric distance measurements lead to potentially increased accuracies. However, for simple interferometric experiments, the periodicity of the signal results is ambiguous up to an absolute distance of a half-wavelength. This ambiguity distance can be extended by combining signals from multiple wavelengths. For instance, the Very Large Telescope Interferometer uses dual-field interferometry to reach nanometric accuracy in a 100~m air-filled delay line~\cite{Delplancke:2008}.

It is also possible to combine both time-of-flight and phase measurement to obtain a better sensitivity and an absolute distance measurement. The ideal tool for this is the phase-stabilized mode-locked femtosecond laser which delivers a frequency comb that can be seen as a perfect source for multi-wavelength interferometry~\cite{Cui:2008,Lamine:2008,Ye:2004}. The problem we address here is the way to use this tool in a complex environment, in order to perform a dispersion free measurement. We will more precisely treat the example of distance measurement in air independent of the variation of physical parameters such as pressure, temperature, CO$_2$ content or humidity.

In a multi-parameter environment, where many physical factors can affect the accuracy of distance measurements, these extra parameters need to be measured and their effects compensated. This is the general strategy of the well-known multicolor schemes~\cite{Bender:1965}, which we introduce in the first part of this paper.

In the second part, we derive fundamental limits for optimal measurement schemes (i.e.\ reaching the Cramér-Rao bound with coherent states~\cite{Helstrom:1968,Braunstein:1994,Pinel:2012})  for distance measurements through a dispersive medium, using mode-locked lasers. The technique is based on temporal mode-dependent interferometry. We show that, in contrast to multicolor schemes for instance, only one measurement is necessary whatever the number of parasitic parameters we want to cancel. This is done at the cost of a precise spectral mode shaping of the frequency comb that is used.

We finally propose a general all-optical experimental setup using pulse shaping and homodyne detections  to reach the previous limit. Our distance measurement can be made insensitive of the environment parameters such as temperature or pressure. No post-processing is needed to achieve a measurement limited by the laser noise. Another advantage over existing schemes is that mode-locked lasers are remarkable tools for optical measurement because of their intrinsic high stability.

\subsection*{Distance measurement in air : multicolor schemes}

Let us start by describing the general technique of multicolor schemes. If one needs to measure a distance in air, the fluctuations of parameters such as pressure or temperature limit the achievable accuracy. The general idea to solve this problem is to perform several measurements at different wavelengths to gain informations about these parameters and compensate the measured length for their variations. For example, in the two-wavelength interferometry (2WI), a given distance $L$ is measured using two wavelengths $\lambda_1$ and $\lambda_2$. The two observables $L_{\phi_1}$ and $L_{\phi_2}$ deduced from the measurement are such that~\cite{Bender:1965,Matsumoto:1992,Meiners:2008}
\begin{equation}
  \label{eq:4}
  L_{\phi_1}=n_\phi(\lambda_1)L\quad\mathrm{and}\quad L_{\phi_2}=n_\phi(\lambda_2)L\;.
\end{equation}
For dry air (pressure of water vapor $P_w=0$), one finds
\begin{equation}
  \label{eq:5}
  L=L_{\phi_1}+\alpha(L_{\phi_1}-L_{\phi_2})\quad\mathrm{with}\quad 
\alpha=\frac{K(\lambda_1)}{K(\lambda_2)-K(\lambda_1)}\;,
\end{equation}
where $K(\lambda)$ is given in Appendix~\ref{sec:airmodel}, and is calculated from the Edlén model of air. The parameter $\alpha$ is independent from pressure, temperature and CO$_2$ content of air. The standard quantum limit for a distance measurement is obtained by taking $(\delta L_{\phi_i})^{\mathrm{shot}}=\frac{c}{2\sqrt{N}\omega_i}$. The sensitivity $\delta L$ on the distance then strongly depends on the value of $\alpha$. Typically, for $\lambda_1=1064\;\mu$m and $\lambda_2=532\;\mu$m, $\alpha$ is of the order of 60. Hence, for a mean photon number $N=4\times10^{16}$ per wavelength ($10$ mW power and an integration time of $1$ s), one gets
\begin{equation}
  \label{eq:21}
   (\delta L) ^\shot_{2WI}\simeq3\times10^{-14}\;\mathrm{m}\;.
\end{equation}
This is the shot noise limit in distance variation measurement (displacement), but one should note that for an absolute distance characterization (ranging), the precise knowledge of $\alpha$ is the main limitation factor ($\delta\alpha/\alpha\sim1\%$).

For moist air, one can no longer use Eq.~(\ref{eq:5}). If the pressure of water vapor is unknown or changing over time, it leads to a systematic error (see~\cite{Meiners:2008} for a discussion). A solution is then to consider a third wavelength
$\lambda_3$~\cite{Golubev:1994} and a third measurement $L_{\phi_3}$ so that
\begin{equation}
  \label{eq:6}
  L=L_{\phi_1}+\beta(L_{\phi_2}-L_{\phi_1})+\gamma(L_{\phi_3}-L_{\phi_1})\;.
\end{equation}
Expressions of $\beta$ and $\gamma$ can be found in~\cite{Golubev:1994} and do not depend on pressure, temperature, CO$_2$ content and humidity. Here again these factors can be large, reducing the sensitivity of displacement measurement. For a same total number of photons and typical wavelengths, the shot noise limit is now around $10^{-12}\;\mathrm{m}$. In addition to this degradation, the three-wavelength scheme is experimentally more involved.

From this example one sees that both the required sensitivity and physical characteristics of the medium conditions on the number of extra parameters one has to take into account. For each of these parameters an additional measurement at a different wavelength is necessary. 

\section{Efficient measurement through dispersive media}

We will now give a more general and systematic approach to this problem. First, we derive general equations for the propagation of light through a dispersive medium whose characteristics depend on external parameters. We then elaborate on the ideas developed in~\cite{Lamine:2008,Pinel:2012} and deal with a very general approach on how to efficiently measure parameters affecting the propagation of a light pulse. We derive fondamental limits imposed by the quantum nature of light. One should note nevertheless that we limit ourselves to the study of coherent states, non-classical quantum states being beyond the scope of this article.

\subsection{Propagation in a dispersive medium}

We consider the propagation of an electromagnetic field along the $z$ direction in a weakly dispersive medium. Its propagation from a source to a detector can be affected by a given set of parameters $\vec p=(p_1,\ldots,p_i,\ldots)$ that modify the propagation distance $L$ and/or the characteristics of the dispersive medium: these may be environmental parameters such as air pressure, temperature, etc., or a physical displacement of the source (or the detector). Neglecting polarization effects and using the paraxial approximation, we assume the field to be in a single transverse mode (such as a TEM$_{00}$ mode), and thus will not write the transverse dependence of the field. We further assume Fourier-limited pulses (assuming perfect temporal coherence) and write the dispersed field, as seen by the detector, as a scalar field:
\begin{equation}
\E(t,\vec p) \equiv \E_0\,u(t,\vec p)\;,
\end{equation}
where $u(t,\vec p)$ is the normalized mean field mode (integrated over the measurement time of the detector) and $\E_0$ is a normalization constant which depends on the number of photons $N$.

In the following, it will be more convenient to work in the Fourier space:
\begin{equation}
\E(\omega,\vec p)\equiv\int\E(t,\vec p)\;\e^{\ii\omega t}\,\dd t, \qquad u(\omega,\vec p)\equiv\int u(t,\vec p)\;\e^{\ii\omega t}\,\dd t\;.
\end{equation}
We define $\omoy$ and $\ovar$ as the mean value and variance of the field:
\begin{equation}
\omoy  =  \int\omega\left\vert u(\omega,\vec p)\right\vert^2\dd\omega,\qquad
\ovar^2 = \int(\omega-\omoy)^2\left\vert u(\omega,\vec p)\right\vert^2\dd\omega\;.
\end{equation}
Let us consider an input field $\E_i(\omega)$ whose frequency profile is known. For the sake of simplicity, the field is considered Gaussian (the same final result can be reached with a non Gaussian field but it involves more complex calculations). This field propagates on a distance $L(\vec p)$ through a dispersive medium with a dispersion relation $k(\omega,\vec p)$ which depends on $\vec p$ through the refractive index $n_\phi(\omega,\vec p)$. In the frequency space, this propagation is characterized by a spectral phase $k(\omega,\vec p)L(\vec p)$:
\begin{equation}
\label{eq:spectralphase}
\E(\omega,p) = \E_i(\omega) \e^{\ii k(\omega,\vec p) L(\vec p)}, \qquad k(\omega,\vec p)=\frac{n_\phi(\omega,\vec p)\omega}{c}\;.
\end{equation}
In this paper, we neglect any absorption of the medium, i.e.~we consider a real refractive index. The previous equation~(\ref{eq:spectralphase}) is therefore also valid by replacing fields $\E$ by normalized modes $u$.

For a weakly dispersive medium, the dispersion relation $k(\omega,\vec p)$ can be expanded to the second order around the mean frequency $\omoy$:
\begin{equation} 
\label{eq:Efourier}
\E(\omega,\vec p) \approx \E_i(\omega)\exp\left[\ii\left(\omoy\, t_\phi(\vec p) + (\omega-\omoy) \,t_g(\vec p) +\frac{(\omega-\omoy)^2}{\omoy} \,t_\GVD(\vec p)\right)\right]\;,
\end{equation}
where 
\begin{eqnarray}
t_\phi(\vec p) & = & n_\phi(\omoy,\vec p)\,\frac{L(\vec p)}{c}\;, \\ 
t_g(\vec p) & = & n_g(\omoy,\vec p)\,\frac{L(\vec p)}{c} = \left(n_\phi(\omoy,\vec p) + \omoy n'_\phi(\omoy,\vec p)\right) \,\frac{L(\vec p)}{c}\;,\\
t_\GVD(\vec p) & = & \omoy\left(n'_\phi(\omoy,\vec p)+\frac{\omoy}{2}n''_\phi(\omoy,\vec p)\right)\, \frac{L(\vec p)}{c}\;. 
\end{eqnarray}
$n_g$ is the group index, and $2\left(n'_\phi(\omoy,\vec p)+\frac{\omoy}{2}n''_\phi(\omoy,\vec p)\right)$ corresponds to the Group Velocity Dispersion (GVD). 

In the temporal domain, the field $\E(t,p)=e^{-i\omega_0 t}\tilde{\E}(t,p)$, the envelope $\tilde{\E}$ of the field travels at the group velocity $\frac{c}{n_g}$ while the carrier $\omega_0$ moves at the phase velocity $\frac{c}{n_\phi}$; a non zero group velocity dispersion leads to a broadening of the envelope.

Any change of a parameter in $\vec p$ that affects the distance $L(\vec p)$ will contaminate all quantities $t_\phi$, $t_g$ and $t_\GVD$, as well as any variation of the refractive index of the medium. In Section~\ref{sec:air}, we show how to uncouple, in air, variation of $L$ from variation due to four different environmental parameters: pressure, temperature, humidity and CO$_2$ content. Note that a generalization to other environmental parameters can be obtained by expanding equation~(\ref{eq:Efourier}) to higher orders of the spectral phase and applying the methods developped later in the paper. Nevertheless, these 4 parameters are typically the most relevant for air.

\subsection{Detection scheme and Cramér-Rao bound}

\label{sec:HDscheme}

\begin{figure}[htpb] 
\begin{center} 
\includegraphics[width=0.6\textwidth]{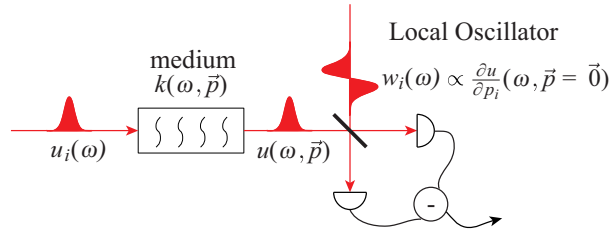} 
\caption{Detection scheme for measuring $p_i$ at the Cramér-Rao bound.}
\label{fig:homodyne}
\end{center} 
\end{figure}

The general problem of estimating a parameter $p_i \in \vec p$ encoded in a light beam $\E(\vec p)$ has been treated in~\cite{Helstrom:1968,Refregier:2004}. The ultimate limit of sensitivity in the measurement of $p_i$ is given by the so-called quantum Cramér-Rao bound, which can be computed once we specify the quantum state of the light beam (coherent state, squeezed state, entangled state etc...)~\cite{Pinel:2012}. For Gaussian states, this Cramér-Rao bound can be reached experimentally with a balanced homodyne detection scheme, as represented in Figure~\ref{fig:homodyne}. The general idea is that the homodyne detection signal is proportional to the projection of the input field into the Local Oscillator (LO) mode.

For a small variation of the set of parameters $\vec{p}$, the field reads:
\begin{equation}
\label{eq:22}
\E(\omega,\vec{p})\approx\E(\omega,\vec{0})+\vec{p}.\overrightarrow\nabla_{\vec{p}}\;\E(\omega,\vec p=\vec 0) = \E_0 \left(u(\omega,\vec{0})+\sum_i p_i \K_i w_i(\omega)\right)\;,
\end{equation}
where $w_i$ are normalized modes such as $w_i(\omega) = \frac{1}{\K_i} \DP{u}{p_i}(\omega,\vec p =\vec 0)$, and $\K_i$ are dimensional normalization constants. Introducing the standard $L^2$ inner product  $\langle f,g\rangle = \int{f^*(\omega) g(\omega)\dd\omega}$, one simply has $\K_i \equiv \sqrt{\langle\DP{u}{p_i}, \DP{u}{p_i}\rangle}$.

One should note that in general the modes $w_i$ do not form an orthogonal basis. This implies that independent measurements of each parameter become a complex problem that we now discuss in detail extensively. Let us first consider the case where only one parameter $p_i \in \vec p$ is influencing the length measurement. It is shown in \cite{Delaubert:2006} that in a homodyne detection scheme, if the LO mode is proportional to $w_i$ (and if there is no phase difference between the LO and the signal $\E$), the detected signal $S[w_i]$ is given by
\begin{equation}
S[w_i] = \frac{1}{\K_i} \mathrm{Re} \left[\langle u(\vec p), w_i\rangle\right] \underset{p_{j\neq i}=0}{=} p_i. 
\end{equation}
For a coherent state illumination with a mean photon number $N$,  the noise in the measurement is $\Delta p_i = 2\sqrt{N}\K_i$. Therefore, the smallest $p_i$ that can be measured (i.e.~for a signal to noise ratio equal to one, $S/\Delta p=1$) is given by
\begin{equation}
(p_i)_{\min} = \frac{1}{2\sqrt{N}\K_i}.
\end{equation}
This value coincides with the Cramér-Rao bound with coherent states~\cite{Pinel:2012}. This shows that a homodyne detection with a LO shaped in mode $w_i$ defines what is called an efficient measurement of $p_i$. Moreover, one sees from the previous expression that the sensitivity of the measurement depends both on the number of photons $N$ and on $\K_i$, the latter reflecting the characteristic variation of the mode with the parameter $p_i$. From now on, the mode $w_i$ will be called the detection mode of the parameter $p_i$.

An experimental demonstration of the efficiency of such a scheme has been realized for parameters corresponding to spatial displacement of a beam~\cite{Hsu:2004, Delaubert:2006}, and a theoretical proposition for a time delay through a dispersion-less medium has been made in~\cite{Lamine:2008}. One should stress that, generally speaking, these kind of experiments are sensitive to variation of parameters within the detection bandwidth, limited here by light time travel. On the other, this system is immune to any fluctuations, whatever their frequencies, of parameter corresponding to orthogonal modes, given we stay in the linear regime.

Let us now consider the general case where there exists at least another parameter $p_{j\neq i}$ such as $w_i$ and $w_j$ are not orthogonal. In that case, a homodyne detection with LO mode $w_i$ will also be sensitive to $p_j$. We show here that this issue can be resolved by 'purifying' the detection mode $w_i$ into a new mode $w_i^p$ that is now orthogonal to $w_j$. This new shape of the LO allows to measure $p_i$ independently of $p_j$. Nevertheless, because it differs from the detection mode $w_i$, it leads to a reduced sensitivity in the measurement of $p_i$. In the general situation, the purified mode for a given parameter $p_i$ is orthogonal to the hyperplane formed by $\{w_{j\neq i}\}$, and the normalization factors are given by $\K_i^p = \K_i \langle w_i^p,w_i\rangle <\K_i$. The sensitivity in the measurement of $p_i$ is therefore decreased and given by $(p_i)_{\min}^p = \frac{1}{2\sqrt{N}\K_i^p}$.

The choice for the mode of the LO is an experimental tradeoff between accuracy (no perturbation coming from $p_j$) and precision. We further elaborate on this point in the following, taking as a simple example the measurement of $t_\phi$, $t_g$ and $t_\GVD$ introduced previously. 

To this aim, we introduce a controlled perturbation $p_\phi \ll t_\phi$ of the phase delay $t_\phi \to t_\phi+p_\phi$, a perturbation $p_g \ll t_g$ of the group delay $t_g\to t_g+p_g$ and a perturbation $p_\GVD \ll t_\GVD$ of the group velocity dispersion delay $t_\GVD \to t_\GVD+p_\GVD$. The corresponding detection modes are given by:
\begin{eqnarray}
w_\phi(\omega) & = & \ii u(\omega) = v_0(\omega) \\
w_g(\omega) & = & \ii \frac{\omega-\omoy}{\ovar} u(\omega) = v_1(\omega) \\
w_\GVD(\omega) & = & \ii \frac{1}{\sqrt{3}}\frac{(\omega-\omoy)^2}{\ovar^2}u(\omega) = \frac{1}{\sqrt 3}v_0(\omega) + \sqrt{\frac{2}{3}} v_2(\omega)
\end{eqnarray}
where we have introduced $\{v_i(\omega)\}$ the orthonormal basis of spectral Hermite-Gaussian modes whose expressions are given in Appendix~\ref{sec:HG}. The normalization factors are given by $\K_\phi = \omoy$, $\K_g = \ovar$ and $\K_\GVD = \sqrt 3 \frac{\ovar^2}{\omoy}$.

It is clear that $w_\phi$ and $w_\GVD$ are not orthogonal, which implies that a measurement using a LO $w_\phi$ will not be accurate because of its sensitivity to $p_\GVD$, and vice versa. More precisely, measurements over the various detection modes will yield the following signals:  
\begin{eqnarray}
S[w_\phi] & = & p_\phi + \frac{\ovar^2}{\omoy^2} p_\GVD \\
S[w_g] & = & p_g \\
S[w_\GVD] & = & \frac{1}{3}\frac{\omoy^2}{\ovar^2} p_\phi + p_\GVD.
\end{eqnarray}
In order to measure only $p_\phi$, one has to consider the \emph{purified mode} $w_\phi^p$ introduced previsously, which is orthogonal to the detection modes of the other parameters. Formally, $w_\phi^p(\omega)$ is orthogonal to the hyperplane generated by $\{w_g,w_\GVD\}$ in the vector space $\{w_\phi,w_g,w_\GVD\}$ and is given in the present case (up to a scalar factor) by:
\begin{equation}
w_\phi^p(\omega) = \sqrt{\frac{2}{3}} v_0(\omega) - \frac{1}{\sqrt 3}v_2(\omega).
\end{equation}
A measurement with LO $w_\phi^p$ yields:
\begin{equation}
S[w_\phi^p] = p_\phi.
\end{equation}
Defining $\K_\phi^p = \K_\phi \langle w_\phi^p, w_\phi\rangle = \sqrt{\frac{2}{3}} \omoy $, the sensitivity of this measurement is:
\begin{equation}
(p_\phi)_{\min}^p = \frac{1}{2\sqrt{N}\sqrt{\frac{2}{3}}\omoy}.
\end{equation}

The two previous equations show that it is possible to retrieve a phase delay information insensitive to any group velocity dispersion fluctuations with only one homodyne measurement. This improvement in accuracy is made at the cost of a decreased precision, determined by the overlap between the purified and non purified modes (here the degradation is given by $\sqrt{3/2}$).

\begin{figure}[htpb] 
\begin{center} 
\includegraphics[width=0.5\textwidth]{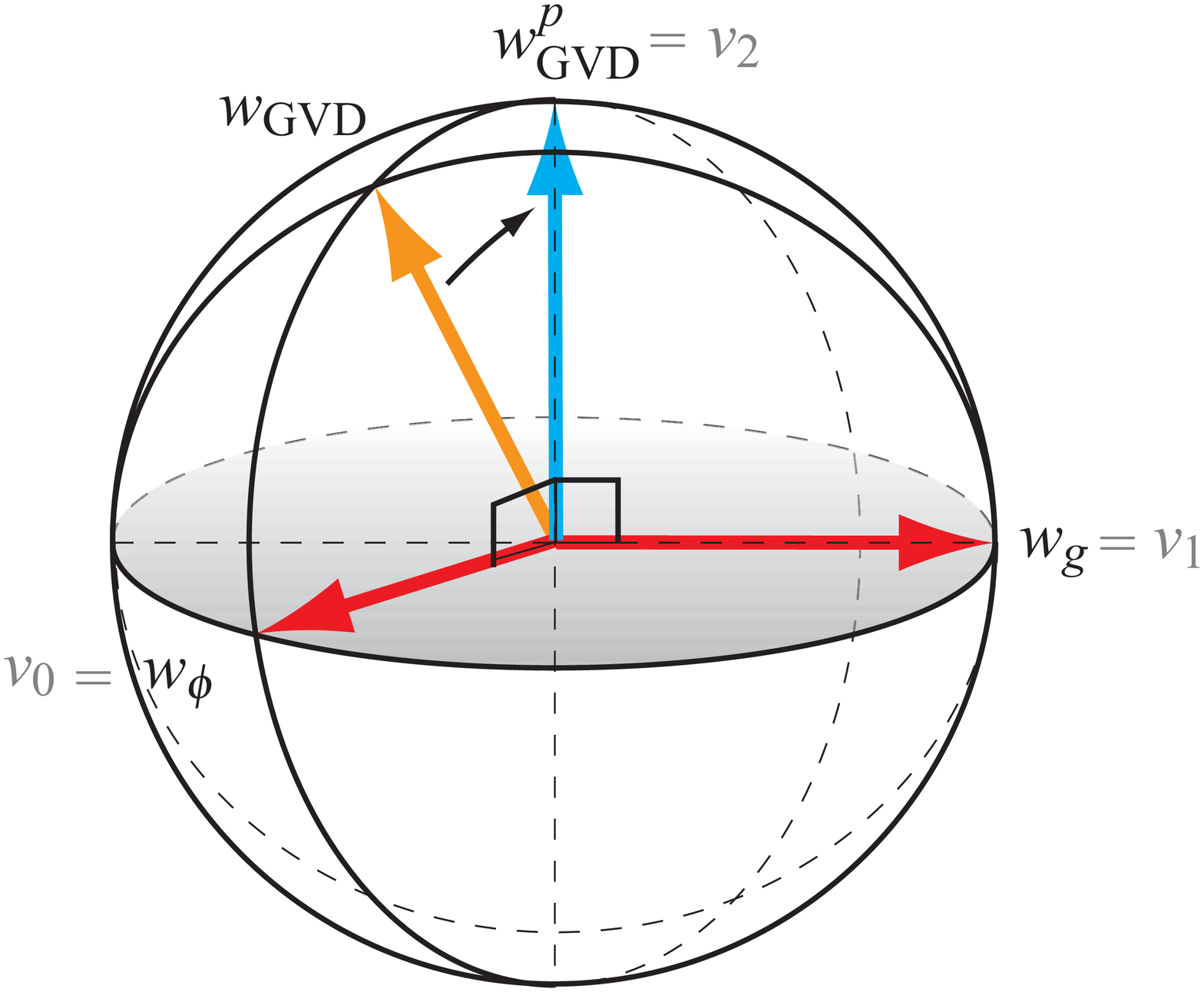}
\includegraphics[width=0.5\textwidth]{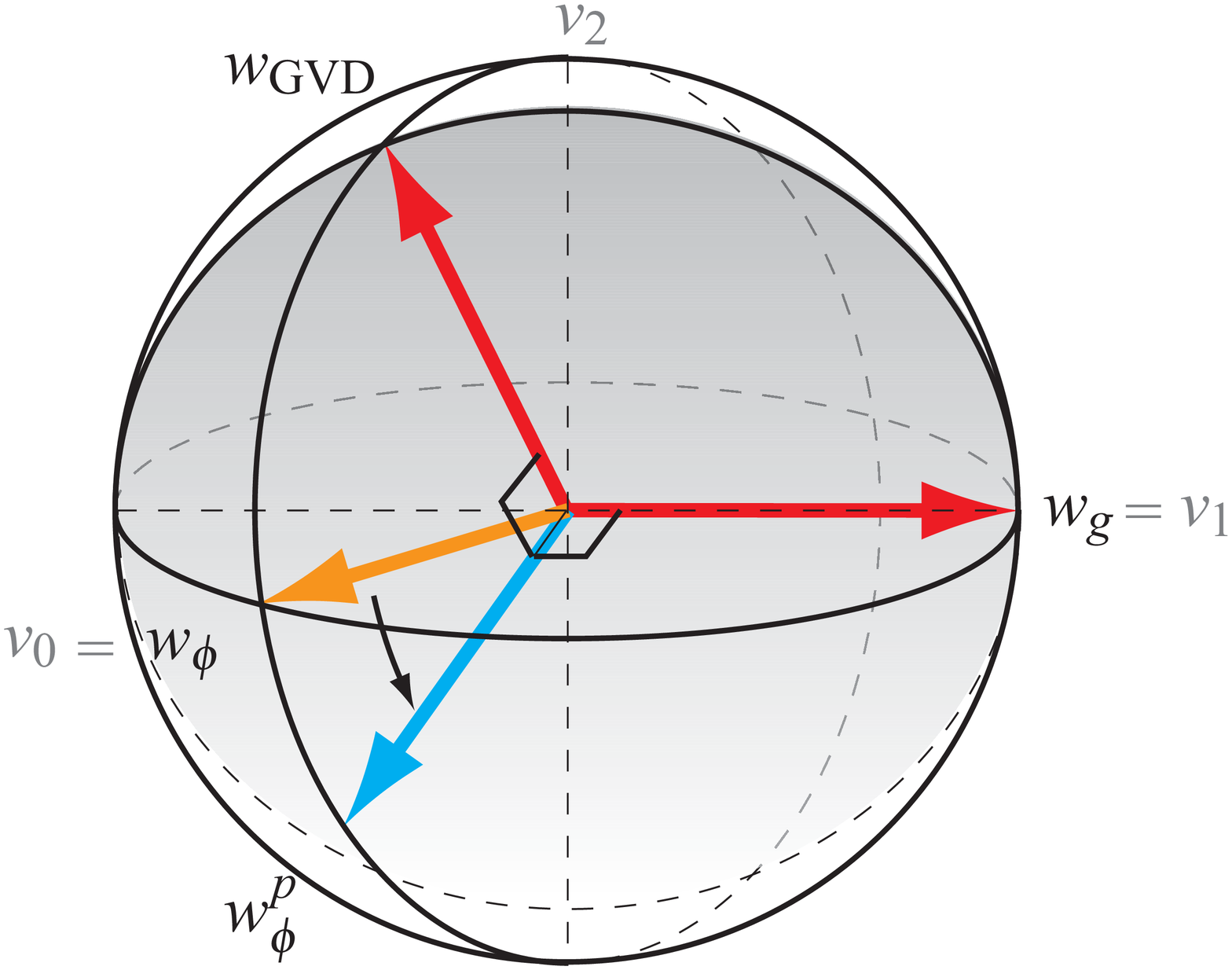}
\caption{Relation between the different LO modes in the vector space $\{v_0,v_1,v_2\}$. The modes $w_\phi^p$, $w_\phi$, $w_\GVD$ and $w_\GVD^p$ lie in the same plane.}
\label{fig:modes_pur}
\end{center} 
\end{figure}

The same process can be applied to obtain the purified mode for measuring $p_\GVD$:
\begin{equation}
w_\GVD^p(\omega) = v_2(\omega) \qquad \mathrm{with} \qquad \K_\GVD^p = \sqrt{2}\frac{\ovar^2}{\omoy}.
\end{equation}
The link between the detection modes and the purified modes is shown in Figure~\ref{fig:modes_pur} and in Table~\ref{table:modes}.

\begin{table}[htpb]
{\renewcommand{\arraystretch}{1.5}
\renewcommand{\tabcolsep}{0.2cm}
\begin{tabular}{|c||c|c|c|}
\hline Parameter & Mode & Sensitivity & Temporal profile\\
\hline
$p_\phi$ & Detection: $w_\phi = v_0$ & $\frac{1}{2\sqrt{N}\omoy}$ & \includegraphics[width=0.15\textwidth]{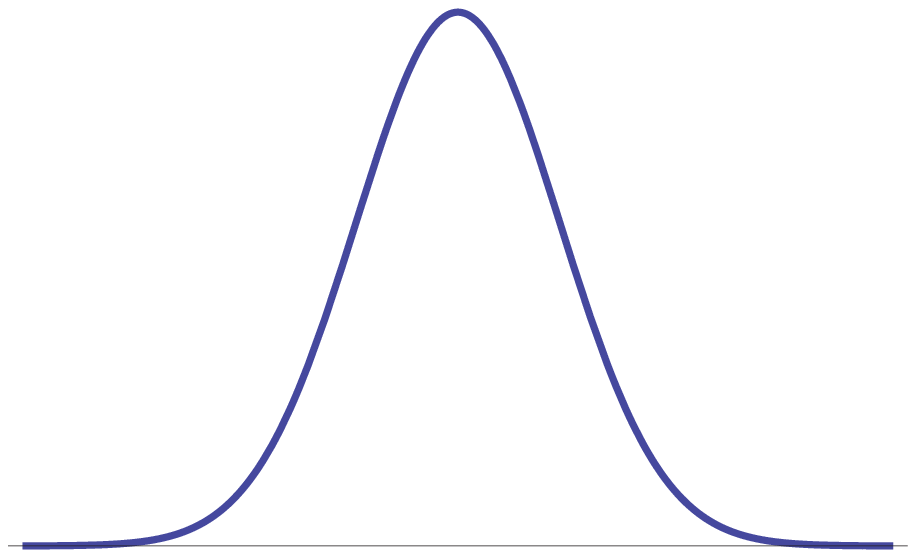} \\ \cline{2-4}
 & Purified: $w_\phi^p = \sqrt{\frac{2}{3}} v_0 - \frac{1}{\sqrt 3} v_2$ & $\frac{1}{2\sqrt{N} \sqrt{\frac{2}{3}}\omoy}$ & \includegraphics[width=0.15\textwidth]{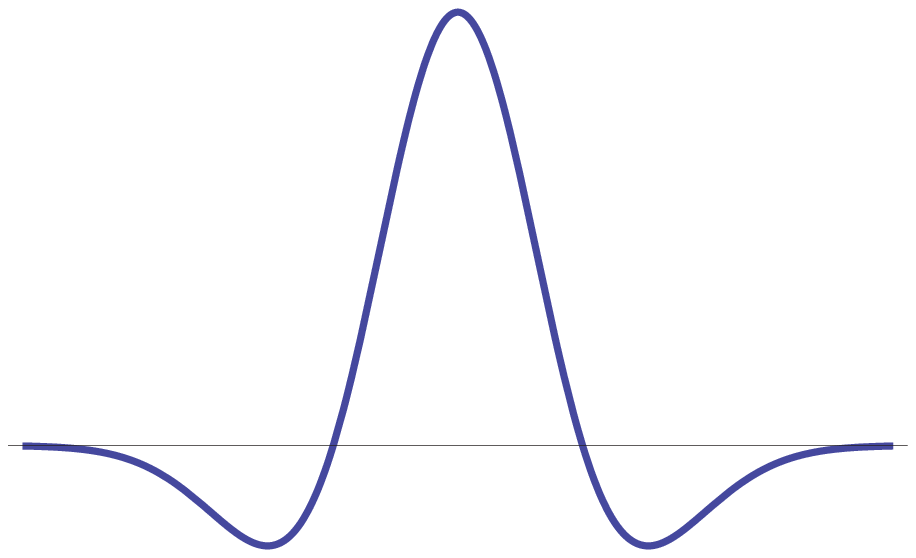} \\
\hline
$p_g$ & $w_g = v_1$ & $\frac{1}{2\sqrt{N}\ovar}$ & \includegraphics[width=0.15\textwidth]{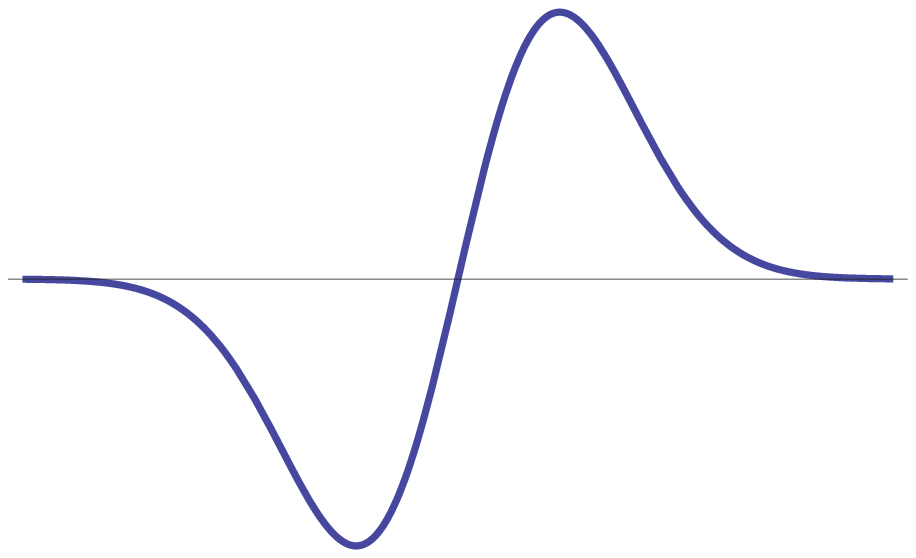} \\
\hline
$p_\GVD$ & Detection: $w_\GVD = \frac{1}{\sqrt 3} v_0 + \sqrt{\frac{2}{3}} v_2$ & $\frac{1}{2\sqrt{N}}\frac{\omoy}{\sqrt{3}\ovar^2}$ & \includegraphics[width=0.15\textwidth]{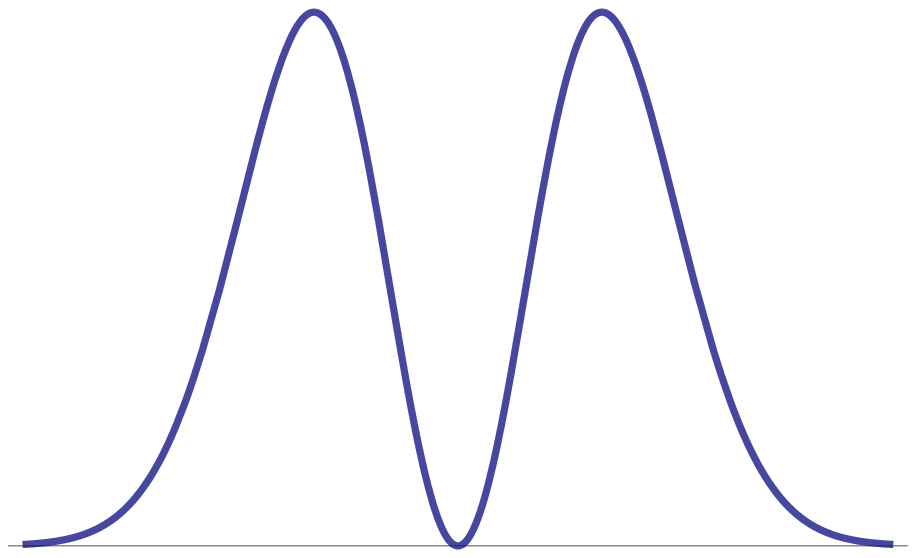} \\ \cline{2-4}
 & Purified: $w_\GVD^p = v_2$ & $\frac{1}{2\sqrt{N}}\frac{\omoy}{\sqrt{2}\ovar^2}$ & \includegraphics[width=0.15\textwidth]{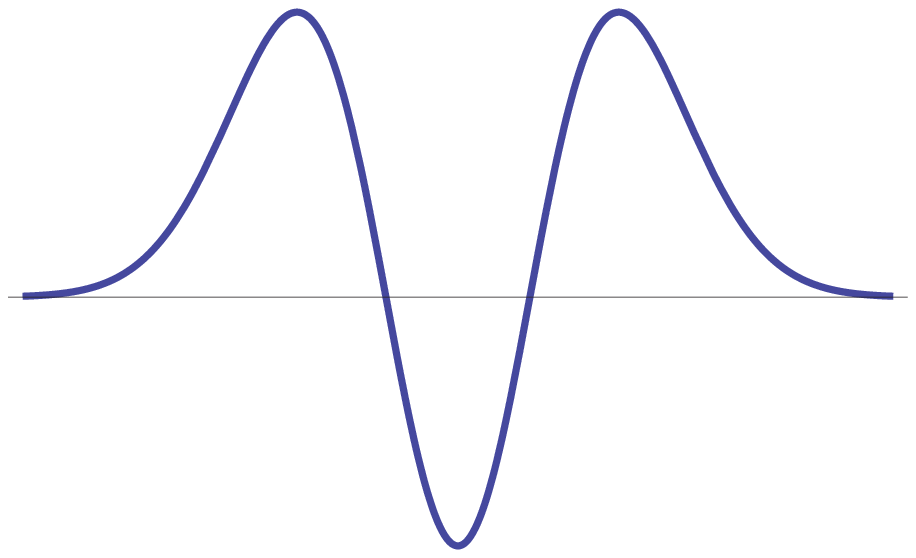} \\
\hline
\end{tabular}}
\caption{Summary of the different LO modes and the sensitivities.}
\label{table:modes}
\end{table}

\section{Application to the measurement of a distance in air}
\label{sec:air}

Let us now consider the specific case of measuring a distance in air,
independently of the fluctuation of environmental parameters such as
pressure and temperature. Indeed, induced index refraction
fluctuations are the main limitations to precise distance
measurement. To access the absolute length, one needs to know
precisely the air index variation with these parameters. One solution is to measure precisely these parameters and use an air model to calculate the refractive index, for instance the Edlén~\cite{Edlen1966} or the Ciddor~\cite{Ciddor:1996} equations. But these techniques are not immune to local variation of the parameters, and some parameters such as the partial pressure of water vapor are very difficult to precisely access. One can also measure directly the local refractive index using a refractometer~\cite{Bonsch:1998}.

Another solution is to make several measurements in order to compensate for these effects. This is the principle of multicolor interferometry~\cite{Bender:1965}, whose state of the art is based on second harmonic generation~\cite{Ishida:1989, Matsumoto:1992}, as we have presented in the introduction. Here we develop our new experimental scheme based on mode-locked lasers interferometry as introduced in the previous section. This technique is another kind of multicolor interferometry and allows for direct measurement of distance in air independently of parameters from the environment.

\subsection{Ranging through air}

We apply the technique developed in Section~\ref{sec:HDscheme} to the measurement of the absolute distance $L$ in air. The parameter to be measured with high sensitivity is $p_L=L$. Fluctuations of the environment do perturb this measurement. They can be separated into two groups of parameters. Firstly temperature $T$, pressure $P$ and CO$_2$ content $x$ affect air index though the same function, as can be seen in the air model developed in Appendix \ref{sec:airmodel}. They will be described by only one parameter $p_X=X(T,P,x)$. Secondly, pressure of water vapor $P_w$ has an independent influence, for which we define the parameter $p_{P_w}=P_w$.

One can calculate the corresponding detection modes using the Edlén model of refractive index recalled in the appendix and the second order development of the electric field introduced in the first section. The distance detection mode is given by :
\begin{equation}
w_L(\omega)  =  \frac{1}{c\K_L} \left(\omoy v_0(\omega) + \ovar \,v_1(\omega)\right)
\end{equation}
and the two other detection modes are given in the appendix.

\begin{figure}[htpb] 
\begin{center} 
\includegraphics[width=0.9\textwidth]{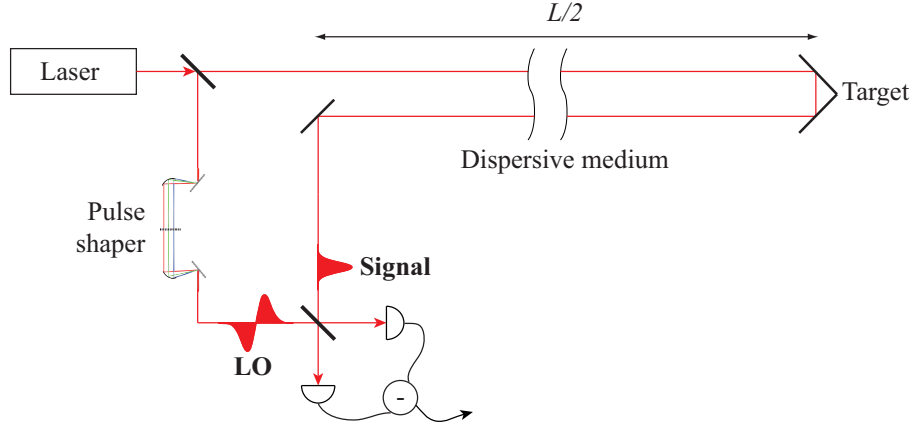} 
\caption{Direct distance measurement with an appropriately shaped LO.}
\label{fig:exp} 
\end{center} 
\end{figure}

These modes are not linearly independent. Thus if we consider a possible experimental scheme with homodyne detection in the detection mode for $L$ (see Figure~\ref{fig:exp})
%\begin{equation}
%t_L = \frac{L}{c}\, , \qquad t_X=K(\omoy) X(T,P,x) \frac{L}{c} \, ,\qquad t_{P_w}=g(\omoy) P_w \frac{L}{c}
%\end{equation}
the measured signal will be~:
\begin{equation} \label{eq:Mranging}
S[w_L]= p_L + \frac{\K_X}{\K_L} \langle w_L,w_X \rangle p_X + \frac{\K_{P_w}}{\K_L} \langle w_L,w_{P_w}\rangle p_{P_w}.
\end{equation}
Therefore, the signal will be contaminated by variations of the different parameters $p_X$ and $p_{P_w}$. Let us first compute the shot noise limit in the case where $p_X$ and $p_{P_w}$ are zero (or sufficiently small). To compare with multicolor schemes, in the remainder of this section the measurement is performed using $N=8\times10^{16}$ photons and assuming a laser bandwidth of $\ovar = \frac{\omoy}{6}$ (corresponding to 3~fs FWHM Fourier-limited pulses). The shot noise limited sensitivity to displacement is about $2\times 10^{-16}\;\mathrm{m}$, comparable to usual interferometric measurement schemes.

One can evaluate the contamination from the other parameters calculating the pre-factor of $p_X$ and $p_{P_w}$ in equation (\ref{eq:Mranging}). One finds $\frac{1}{L}\frac{\K_X}{\K_L} \langle w_L,w_X \rangle = 27\times 10^{-5}\;\mathrm{Pa}^{-1}$ and $\frac{1}{L}\frac{\K_{P_w}}{\K_L} \langle w_L,w_{P_w}\rangle =  -3.7\times 10^{-10}\;\mathrm{Pa}^{-1}$. These factors are big compared to shot noise limited pure distance measurements. It means it is necessary in this case to take into account the parasitic parameters in order to preserve the accuracy of the measurement. To solve this issue, we can apply the detection mode purification procedure introduced previously~:
\begin{eqnarray}
w_L^p(\omega) \propto w_L(\omega) & - & \frac{\langle w_L,w_{P_w}\rangle - \langle w_X,w_{P_w}\rangle \langle w_L,w_X\rangle}{1-\langle w_X,w_{P_w}\rangle^2} w_{P_w}(\omega)\\
 & - & \frac{\langle w_L,w_X\rangle - \langle w_X,w_{P_w}\rangle \langle w_L,w_{P_w}\rangle}{1-\langle w_X,w_{P_w}\rangle^2} w_X(\omega)\;.
\end{eqnarray}
The spectral profiles of the purified and non-purified modes are plotted in Figure \ref{fig:modecompare}. The normalization constant as well as the derivation of this mode can be found in Appendix~\ref{sec:detectionair}. It is found that~:
\begin{equation}
\K_L^p = \K_L \sqrt{1-\frac{\langle w_L,w_{P_w}\rangle^2 + \langle w_L,w_X\rangle^2-2 \langle w_L,w_{P_w}\rangle \langle w_L,w_X\rangle\langle w_X,w_{P_w}\rangle}{1-\langle w_X,w_{P_w}\rangle^2}}\;.
\end{equation}
In that case, the shot noise limit has the following value~:
\begin{equation}
(\delta L)^\shot_{1\mathrm{HD}}=\frac{c}{2\sqrt{N}\K_L^p} = 2\times10^{-11} \mathrm{m}\;.
\end{equation}
As a matter of comparison, a purification for $p_X$ only (which would be equivalent to the two-color scheme) leads to a shot noise limit of $(\delta L)^\shot_{1\mathrm{HD}}=\frac{c}{2\sqrt{N}\K_L^p} = 3\times10^{-13} \mathrm{m}$. Even if the precision is 2 orders of magnitude better, the accuracy of the measurement is not controlled because of the unknown value of humidity introducing a systematic error.

\begin{figure}[htpb] 
\begin{center} 
\includegraphics[width=0.7\textwidth]{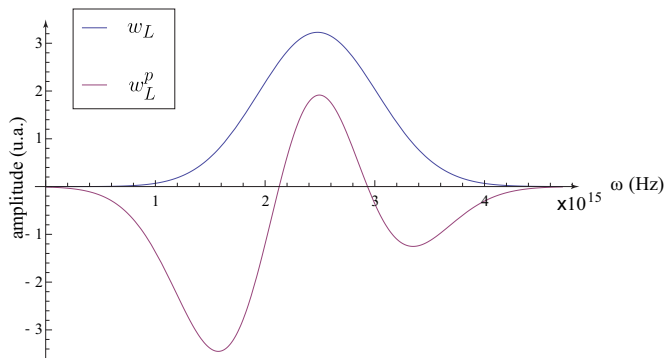} 
\caption{Spectral profile of modes $w_L$ and $w_L^p$ for 3~fs FWHM Gaussian pulses.}
\label{fig:modecompare} 
\end{center} 
\end{figure}

Using that scheme, one can perform a single-shot measurement of distance immune from environmental parameters fluctuations. Sensitivity is of the same order of magnitude than in multicolor scheme (slightly lower in our examples, but this is simply due to smaller spectral width), and depends on how many parameters one wants to get immune from. From a more technical point of view, no complex frequency generation is necessary, a femtosecond source is enough to perform the measurement. Furthermore, the scheme described here could be extended to more parameters. Of course, the mode shaping can be more difficult to produce and in particular the precision required in shaping becomes more stringent when more parameters are considered. In a realistic implementation, one can set up active pulse-shaping with search algorithms to determine the proper shaping. It is significant to notice that all the work is to be done at the detection stage and not on the light sent through the medium, which makes it much easier to handle.

\section{Conclusion}

We have exhibited a novel optimal and all-optical scheme to measure in real time a distance compensated for refractive index changes. It relies on a homodyne detection whose local oscillator projects the measurement on an appropriate mode, hence no post-processing is necessary. We believe this is a simplification compared to existing schemes such as spectral interferometry for example, where derivatives of the spectral phase have to be done after the measurement.

Let us finally mention that this scheme can further be improved in
order to go below the standard quantum limit. It is well known that the sensitivity of a measurement can outreach the shot noise limit by using quantum resources such as squeezed or entangled light~\cite{Giovannetti:2004}. In the scheme presented in this paper, this can be achieved by using a multimode signal light beam with squeezing in the detection mode associated to the measurement, as demonstrated in Ref.~\cite{Pinel:2012}.

\appendix

\section{Index of refraction of air}

\label{sec:airmodel}

The common equations used to derive the wavelength dependence of the refractive index of air are given by the Edlén equations~\cite{Edlen1966}, modified by different authors since that time~\cite{Ciddor:1996,Owens:1967,Peck:1972,Birch:1993,Birch:1994,Bonsch:1998}. The accuracy of those equations are roughly of the order of a few $10^{-9}$ for dry air and $10^{-8}$ for moist air and needs to take into account a large number of parameters, usually temperature, pressure, CO$_2$ content, pressure of water vapor. Moreover, different studies do not necessarily agree and do not cover the whole spectrum. A recent precise measurement of the refractive index of air has been done in~\cite{Macovez:2009}. Experimentally, measuring all those parameters can be difficult for certain situations, and in addition those parameters have to be measured in real time, in order to compensate for fluctuations of the refractive index.

In this paper we consider the updated Edlén formula of Bönsch and Potulski~\cite{Bonsch:1998}~:
\begin{equation}
  \label{eq:2}
  n_{\phi}(\sigma,T,P,x,P_w)-1=K(\sigma)\,X(T,P,x)-g(\sigma)\,P_w
\end{equation}
where $\sigma=1/\lambda$ is the wavenumber and
\begin{eqnarray}
\label{eq:3}
  K(\sigma)&=&10^{-8}\left(A+\frac{B}{130-\sigma^2}+
\frac{C}{38.9-\sigma^2}\right)\;,\\
X(T,P,x)&=&\frac{P}{D}\;\frac{1+10^{-8}(E-F\,T)P}{1+G\, T}\left[1+H(x-0.04\%)\right]\;,\\
g(\sigma)&=&10^{-10}\left(I-J\sigma^2\right) \;,
\end{eqnarray}
with $\sigma$ in $\mu$m$^{-1}$, $P$ and $P_w$ in Pascal (Pa), $T$ in degree Celsius ($^\circ C$) and $x$, the CO$_2$ content, in percentage. The different coefficients further read
\begin{align}
  \label{eq:17}
&A=8091.37\quad,\quad B=2333983\quad,\quad C=15518 \;,\\
&D=932164.60\quad,\quad E=0.5953\quad,\quad F=0.009876\quad,\quad
G=0.0036610 \;, \\
&H=0.5327 \;,\\
&I=3.802\quad,\quad J=0.0384\;.
\end{align}

The group index is therefore given by
\begin{equation}
  \label{eq:12}
  n_g(\sigma,T,P,x,P_w)-1=(K(\sigma)+\sigma K'(\sigma)) \,X(T,P,x)-(g(\sigma)+\sigma g'(\sigma)) \,P_w \;,
\end{equation}
so that $n_g-n_\phi=\sigma\left(K'(\sigma)X(T,P,x)-g'(\sigma)P_w\right)$.

\section{Hermite-Gaussian set of spectral modes}

\label{sec:HG}

For a Gaussian mean field mode
\begin{equation}
u(\omega)=\frac{1}{\sqrt{\ovar}}\frac{1}{\left(2\pi\right)^{1/4}}\e^{-\frac{\left(\omega-\omoy\right)^{2}}{4\ovar^{2}}}\;,
\end{equation}
we introduce the Hermite-Gauss modes
\begin{equation}
v_{n}(\omega)=\ii\frac{1}{\sqrt{2^{n}n!}}H_{n}\left(\frac{\omega-\omoy}{\sqrt{2}\ovar}\right)u(\omega).
\end{equation}
These modes $\{v_n(\omega)\}$ form an orthonormal basis of modes.

\begin{figure}[htpb] 
\begin{center} 
\includegraphics[width=0.7\textwidth]{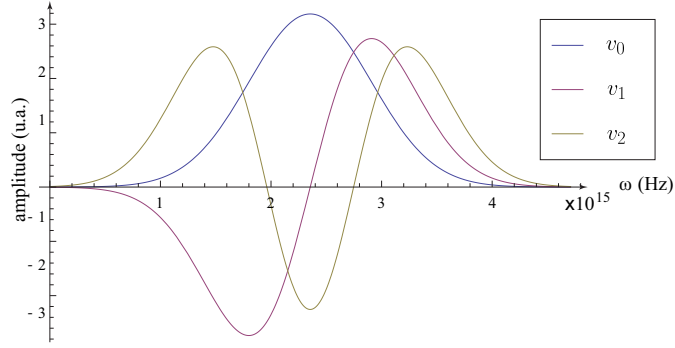} 
\caption{Spectral profile of Hermite-Gaussian modes $v_0$, $v_1$ and $v_2$.}
\label{fig:modeshg} 
\end{center} 
\end{figure}

In order to describe the spectral phase up to the $k^\mathrm{th}$ order, one needs to use the basis $\{v_n(\omega)\}$ up to the mode $k$. In this paper, we develop the spectral phase to the second order; therefore, we use the following modes:
\begin{eqnarray}
v_{0}(\omega) & = & \ii\, u(\omega)\\
v_{1}(\omega) & = & \ii\frac{\omega-\omoy}{\ovar}\, u(\omega)\\
v_{2}(\omega) & = & \ii\frac{1}{\sqrt{2}}\left(\frac{(\omega-\omoy)^{2}}{\ovar^{2}}-1\right)u(\omega).
\end{eqnarray}
The mode spectral profiles are plotted in Figure~\ref{fig:modeshg}.

\section{Detection modes of environmental parameters}

\label{sec:detectionair}

Detection modes read :
\begin{eqnarray}
w_L(\omega) & = & \frac{1}{c\K_L} \left(\omoy v_0(\omega) + \ovar v_1(\omega)\right) \\
w_X(\omega) & = & \frac{LK(\omega_0)}{c\K_X} \left[\left(\omoy + \frac{\ovar^2}{\omoy}(\delta_1+\delta_2)\right)v_0(\omega) + \ovar(1+\delta_1)v_1(\omega) + \sqrt 2 \frac{\ovar^2}{\omoy}(\delta_1+\delta_2)v_2(\omega)\right] \nonumber \\ 
w_{P_w}(\omega) & = & \frac{-Lg(\omega_0)}{c\K_{P_w}} \left[\left(\omoy + \frac{\ovar^2}{\omoy}(\eta_1+\eta_2)\right)v_0(\omega) + \ovar(1+\eta_1)v_1(\omega) + \sqrt 2 \frac{\ovar^2}{\omoy}(\eta_1+\eta_2)v_2(\omega)\right] \nonumber
\end{eqnarray}
where we have defined characteristic quantities that do not depend on the environmental parameters $T$, $P$, $x$ and $P_w$:
 \begin{eqnarray}
\delta_1 = \omoy \frac{K'(\omoy)}{K(\omoy)} \, , &\qquad &\delta_2 = \frac{\omoy^2}{2} \frac{K''(\omoy)}{K(\omoy)}\, , \\
\eta_1 = \omoy \frac{g'(\omoy)}{g(\omoy)} \, , &\qquad &\eta_2 = \frac{\omoy^2}{2} \frac{g''(\omoy)}{g(\omoy)}\,.
\end{eqnarray}

Defining :
\begin{eqnarray}
\K_L & = & \frac{1}{c} \sqrt{\omoy^{2}+\ovar^{2}} \\
\K_X & = & \frac{K(\omoy) L}{c} \sqrt{\left(\omoy+\frac{\ovar^{2}}{\omoy}(\delta_{1}+\delta_{2})\right)^{2}+\ovar^{2}(1+\delta_{1})^{2}+2\frac{\ovar^{4}}{\omoy^{2}}(\delta_{1}+\delta_{2})^{2}} \\
\K_{P_w} & = & \frac{g(\omoy) L}{c} \sqrt{\left(\omoy+\frac{\ovar^{2}}{\omoy}(\eta_{1}+\eta_{2})\right)^{2}+\ovar^{2}(1+\eta_{1})^{2}+2\frac{\ovar^{4}}{\omoy^{2}}(\eta_{1}+\eta_{2})^{2}}
\end{eqnarray}

Measurements with the detection modes give:
\begin{eqnarray}
M[w_L] & = & p_L + \frac{\K_X}{\K_L} \langle w_L,w_X\rangle p_X + \frac{\K_{P_w}}{\K_L} \langle w_L,w_{P_w}\rangle p_{P_w} \\
M[w_X] & = & \frac{\K_L}{\K_X}\langle w_L,w_X\rangle p_L+ p_X  + \frac{\K_{P_w}}{\K_X} \langle w_L,w_{P_w}\rangle p_{P_w} \\
M[w_{P_w}] & = & \frac{\K_L}{\K_{P_w}} \langle w_L,w_{P_w}\rangle p_L+ \frac{\K_X}{\K_{P_w}} \langle w_X,w_{P_w}\rangle p_X  + p_{P_w}
\end{eqnarray}

Calculation of the purified mode $w_L^p$: we first calculate a orthonormal basis in the plane $\{w_X,w_{P_w}\}$, which leads for example to a basis $\{w_X,w_{P_w}^i\}$ where
\begin{eqnarray}
w_{P_w}^i(\omega) = \frac{1}{\sqrt{1-\langle w_X,w_{P_w}\rangle^2}}\left(w_{P_w}(\omega)-\langle w_X,w_{P_w}\rangle w_X\right)
\end{eqnarray}
Then we do a Gram-Schmidt orthonormalization of $\{w_X,w_{P_w}^i,w_L\}$ which gives:
\begin{eqnarray}
w_L^p(\omega) \propto w_L(\omega) - \langle w_L,w_X\rangle w_X(\omega) - \langle w_L,w_{P_w}^i\rangle w_{P_w}^i
\end{eqnarray}
The normalization constant is:
\begin{equation}
\sqrt{1-\frac{\langle w_L,w_{P_w}\rangle^2 + \langle w_L,w_X\rangle^2-2 \langle w_L,w_{P_w}\rangle \langle w_L,w_X\rangle\langle w_X,w_{P_w}\rangle}{1-\langle w_X,w_{P_w}\rangle^2}}
\end{equation}

\section*{Acknowledgments}
The research is supported by ANR project Qualitime, ERC starting grant Frecquam and by the Australian Research Council Centre of Excellence for Quantum Computation and Communication Technology, project number CE110001027 (OP).


\begin{thebibliography}{10}
\newcommand{\enquote}[1]{``#1''}

\bibitem{Weiss:2001}
A.~Weiss, M.~Hennes, and M.~Rotach, \enquote{Derivation of refractive index and
  temperature gradients from optical scintillometry to correct atmospherically
  induced errors for highly precise geodetic measurements,} Surveys in
  geophysics \textbf{22}, 589--596 (2001).

\bibitem{Exertier:2006}
P.~Exertier, P.~Bonnefond, F.~Deleflie, F.~Barlier, M.~Kasser, R.~Biancale, and
  Y.~MÈnard, \enquote{Contribution of laser ranging to earth's sciences,}
  Comptes Rendus Geoscience \textbf{338}, 958 -- 967 (2006).

\bibitem{Djerroud:2010}
K.~Djerroud, O.~Acef, A.~Clairon, P.~Lemonde, C.~N. Man, E.~Samain, and
  P.~Wolf, \enquote{Coherent optical link through the turbulent atmosphere,}
  Optics Letters \textbf{35}, 1479--1481 (2010).

\bibitem{Delplancke:2008}
F.~Delplancke, \enquote{The prima facility phase-referenced imaging and
  micro-arcsecond astrometry,} New Astronomy Reviews \textbf{52}, 199 -- 207
  (2008).

\bibitem{Cui:2008}
M.~Cui, R.~N. Schouten, N.~Bhattacharya, and S.~A. Berg, \enquote{Experimental
  demonstration of distance measurement with a femtosecond frequency comb
  laser,} Journal of European Optical Society Rapid Publications \textbf{3},
  08003 (2008).

\bibitem{Lamine:2008}
B.~Lamine, C.~Fabre, and N.~Treps, \enquote{Quantum improvement of time
  transfer between remote clocks,} Physical Review Letters \textbf{101}, 123601
  (2008).

\bibitem{Ye:2004}
J.~Ye, \enquote{Absolute measurement of a long, arbitrary distance to less than
  an optical fringe,} Optics letters \textbf{29}, 1153--1155 (2004).

\bibitem{Bender:1965}
P.~L. Bender and J.~C. Owen, \enquote{Correction of optical distance
  measurement for the fluctuating atmospheric index of refraction,} Journal of
  geophysical research \textbf{70}, 2461--2462 (1965).

\bibitem{Helstrom:1968}
C.~Helstrom, \enquote{The minimum variance of estimates in quantum signal
  detection,} Information Theory, IEEE Transactions on \textbf{14}, 234 -- 242
  (1968).

\bibitem{Braunstein:1994}
S.~Braunstein and C.~Caves, \enquote{Statistical distance and the geometry of
  quantum states,} Physical Review Letters \textbf{72}, 3439--3443 (1994).

\bibitem{Pinel:2012}
O.~Pinel, J.~Fade, D.~Braun, P.~Jian, N.~Treps, and C.~Fabre,
  \enquote{{Ultimate sensitivity of precision measurements with intense
  Gaussian quantum light: A multimodal approach},} Physical Review A
  \textbf{85}, 1--4 (2012).

\bibitem{Matsumoto:1992}
H.~{Matsumoto} and T.~{Honda}, \enquote{{High-accuracy length-measuring
  interferometer using the two-colour method of compensating for the refractive
  index of air},} Measurement Science and Technology \textbf{3}, 1084--1086
  (1992).

\bibitem{Meiners:2008}
K.~{Meiners-Hagen} and A.~{Abou-Zeid}, \enquote{{Refractive index determination
  in length measurement by two-colour interferometry},} Measurement Science and
  Technology \textbf{19}, 084004--+ (2008).

\bibitem{Golubev:1994}
A.~N. Golubev and A.~M. Chekhovsky, \enquote{Three-color optical range
  finding,} Applied Optics \textbf{33}, 7511--7517 (1994).

\bibitem{Refregier:2004}
P.~R{\'e}fr{\'e}gier, \emph{Noise theory and application to physics: from
  fluctuations to information} (Springer Verlag, 2004).

\bibitem{Delaubert:2006}
V.~Delaubert, N.~Treps, C.~Harb, P.~Lam, and H.~Bachor, \enquote{Quantum
  measurements of spatial conjugate variables: displacement and tilt of a
  gaussian beam,} Optics letters \textbf{31}, 1537--1539 (2006).

\bibitem{Hsu:2004}
M.~Hsu, V.~Delaubert, P.~Lam, and W.~Bowen, \enquote{Optimal optical
  measurement of small displacements,} Journal of Optics B: Quantum and
  Semiclassical Optics \textbf{6}, 495 (2004).

\bibitem{Edlen1966}
B.~{Edl{\'e}n}, \enquote{{The Refractive Index of Air},} Metrologia \textbf{2},
  71--80 (1966).

\bibitem{Ciddor:1996}
P.~E. {Ciddor}, \enquote{{Refractive index of air: new equations for the
  visible and near infrared},} Applied Optics \textbf{35}, 1566--+ (1996).

\bibitem{Bonsch:1998}
G.~{B{\"o}nsch} and E.~{Potulski}, \enquote{{Measurement of the refractive
  index of air and comparison with modified Edl{\'e}n's formulae},} Metrologia
  \textbf{35}, 133--+ (1998).

\bibitem{Ishida:1989}
A.~{Ishida}, \enquote{{Two-wavelength displacement-measuring interferometer
  using second-harmonic light to eliminate air-turbulence-induced errors},}
  Japanese Journal of Applied Physics \textbf{28}, L473--L475 (1989).

\bibitem{Giovannetti:2004}
V.~Giovannetti, S.~Lloyd, and L.~Maccone, \enquote{Quantum-enhanced
  measurements: beating the standard quantum limit,} Science \textbf{306}, 1330
  (2004).

\bibitem{Owens:1967}
J.~C. {Owens}, \enquote{{Optical refractive index of air: dependence on
  pressure, temperature, and composition},} Applied Optics \textbf{6}, 51--+
  (1967).

\bibitem{Peck:1972}
E.~R. {Peck} and K.~{Reeder}, \enquote{{Dispersion of Air},} Journal of the
  Optical Society of America (1917-1983) \textbf{62}, 958--+ (1972).

\bibitem{Birch:1993}
K.~P. {Birch} and M.~J. {Downs}, \enquote{{An Updated Edl{\'e}n Equation for
  the Refractive Index of Air},} Metrologia \textbf{30}, 155--162 (1993).

\bibitem{Birch:1994}
K.~P. {Birch} and M.~J. {Downs}, \enquote{{LETTER TO THE EDITOR: Correction to
  the Updated Edl{\'e}n Equation for the Refractive Index of Air},} Metrologia
  \textbf{31}, 315--316 (1994).

\bibitem{Macovez:2009}
R.~Macovez, M.~Mariano, S.~D. Finizio, and J.~Martorell, \enquote{Measurement
  of the dispersion of air and of refractive index anomalies by
  wavelength-dependent nonlinear interferometry,} Optics Express \textbf{17},
  13881--13888 (2009).

\end{thebibliography}
\end{document}